\documentclass[10pt]{article}

\usepackage{psfrag,graphicx,times}
\usepackage{amsfonts,amsmath,amssymb,graphicx,epsfig}

\font\ftitre=ptmb at 14pt
\font\fauthor=ptmb at 8pt
\font\fabstract=ptmr at 9pt

\font\fsection=ptmb at 10pt
\font\soussection=ptmri at 10pt

\def\uneligne{\vspace{14 pt}}
\def\unelignetreize{\vspace{13 pt}}

\makeatletter
\renewcommand{\section}{\@startsection
{section}{0}{0mm}{\baselineskip}{0.3\baselineskip}{\fsection}}
\renewcommand{\subsection}{\@startsection
{subsection}{1}{0mm}{2\baselineskip}{\baselineskip}{\soussection}}
\renewcommand{\subsubsection}{\@startsection
{subsubsection}{2}{0mm}{\baselineskip}{\baselineskip}{\itshape}}
\makeatother 

\begin{document}

\vspace{1 cm}

\begin{center}
  {\ftitre  STATISTICAL MECHANICS OF
 INTERACTING FIBER BUNDLES }\\
\uneligne
{\fauthor Renaud Toussaint\\
Department of Physics, University of Oslo, P.O.Box 1048 Blindern, N-0316 Oslo, Norway.
}
\end{center}

\unelignetreize

{\centerline { \fabstract \bf ABSTRACT}}

{\fabstract  We consider quasistatic fiber bundle models with weak interactions, 
i.e. where the perturbation emanating from every broken fiber is small compared to the mean-field 
 imposed average deformation of the bundle. Classical load sharing rules are considered,
 namely purely local, purely global or decaying as a power-law of distance. All fibers have identical spring 
constants, reducing to zero after their irreversible break, which happens at a random threshold picked ab initio independently 
for every fiber from a quenched disorder (q.d.) distribution. Initially,
 all fibers are intact and as the buffer plates are progressively separated, with a controlled displacement between them, fibers
 break one after the other. We are interested in the probability distribution of configurations of broken fibers, averaged over
 all possible realizations of the underlying q.d. (i.e. over all possible values of the 
set of threshold distributions). This configurational distribution is accessed via integration over the 
independent variables of the system, i.e. through mapping the threshold set space onto the configurational space,
 via paths corresponding to the deterministic evolution of bundles characterized by each set of realized q.d.,
 up to a certain imposed elongation. Using a perturbational approach allows to obtain this configurational distribution
 exactly to leading order in the interactions. 
This maps this fiber bundle systems onto classical statistical mechanics models,
 namely percolation, standard Ising models or generalized Ising models depending on the range of the interactions chosen in the 
load sharing rule. This relates unambiguously such 
q.d. based systems to standard classical mechanics, which allows the use of the associated toolbox to derive various
 observables of the system, as e.g. correlation lengths. The  thermodynamic parameters formally equivalent to temperature and 
chemical potential, are analytically expressed functions of the externally imposed deformation, with functional dependences 
depending on the load sharing rule and the particular choice of the q.d. distribution.
}
\unelignetreize

{\center {\section {INTRODUCTION}}}
The physical process of brittle failure under external load has long been studied, and is well understood in the case of
 a homogeneous solid (Griffith [1] ), but the behavior of mechanically heterogeneous systems is still an open subject of research.
 The difficulty arises from the necessity of quantifying the effect of randomness in the mechanical properties of many 
interacting constituents. Despite many advances over the last 20 years [2], no analytical unified description of such breakdown 
processes of heterogeneous materials is available at the moment.  
Most results in this field are obtained from lattice models, e.g., 
spring or beam networks, or fuse networks, scalar analog of the elastic problem [2].
 These simulations led to an understanding of the experimentally well-established Hurst exponent of fracture surfaces [3]:
 0.8 in the case of three dimensional fracture, with a cross-over to 0.5 at small scales (see Bouchaud for a review [4]),
 or 0.6 for the roughness of a fracture front in interfacial brittle failure in mode I (Schmittbuhl and M{\aa}l{\o}y [5] ). 
Fiber bundle models, first introduced by Daniels [6] and Coleman [7],
 are among the most studied paradigms of simplified lattice models of breakdown processes in heterogeneous materials.
  They consist of a bundle of parallel fibers set under tension between two buffer plates, with random elongation thresholds for breaking,
 and a model-dependent load sharing rule. This rule states how the load carried by a fiber is redistributed when it breaks 
among the surviving fibers, and reflects the physical properties of the buffer plates: purely rigid, elastic, or more complicated. 
The most commonly considered rules are the Global Sharing Rule (GLS), where the load is uniformly distributed among all fibers,
 and the Local Sharing Rule (LLS), where broken fibers only overload the nearest surviving fibers. 
Analytical solutions are available in these two extreme cases for the average load curve (Sornette [8] for GLS) or the statistics of avalanches (Hemmer, Hansen and Kloster [9,10,11]).
 For more general load sharing rules, like these corresponding to plates responding elastically or 
transferring the load as a power-law of distance to broken fibers, only numerical solutions are available [12]. 

In the present paper, we will present a formal analytical mapping of such quasistatic q.d. based models onto standard 
statistical mechanics models, namely percolation, Ising and generalized Ising models,
 depending on the particular disorder distribution and load sharing rule adopted. Specifically,
 we will consider any possible particular (initial) realization of the q.d. describing the set of breaking thresholds of the fibers,
 and compute to which configuration of broken and intact fibers each realization 
leads when the initially intact system is monotonically extended 
from zero to a given fixed extension. Considering then the ensemble of possible realizations of the q.d.,
 we will obtain the probability distribution of the possible damage configurations at the considered extension,
 as the frequency of occurrence of each configuration among all possibles, averaged over all realizations of the q.d..
 Insodoing, we will show that in the limit of small interactions, 
the emerging configurational distribution can be expressed as Boltzmannians of a simple functional of the damage field, 
and relate these distribution to standard statistical mechanics.
 The equivalent of temperature, which will be here a probabilistic energy scale,
 and external field setting the average fraction of broken fibers,
 will be obtained analytically from the underlying q.d., load sharing rule and extension achieved. 

Relating analytically the well known Fiber Bundle Models to such classical models 
of statistical mechanics is important in the sense that it allows to use the traditional toolbox 
of standard statistical mechanics, and possibly 
to classify the possible transitions corresponding to localization of disorder and/or macroscopic rupture.
 Indeed the classification of rupture processes as second or first order transitions, or spinodal nucleation processes,
 is still a subject of debate.
 The difficulty of such classification in this problem lies in the absence of analytical form for the probability 
distribution of configurations of broken elements, apart from the simplest cases (GLS,LLS).
Although such probability distributions in similar systems have been proposed [13], they were in general postulated,
 whereas we will here derive this distribution directly from first principles of evolution rules, 
incorporating the choice of a load sharing rule and threshold distribution.

We also underline that we will only consider quasistatic models, 
in which the disorder is quenched ab initio,
 and in which there is no evolution of microstate when the external parameter, imposed average elongation, is kept fixed. 
This should be relevant to describe systems where the inverse imposed strain rate is significantly lower than any characteristic
 time for thermal transition from a fracture state to another at fixed external elongation, solely due to molecular motion 
(otherwise thermal rupture models with quenched disorder should be considered, which is described in the GLS case e.g. by Politi [14]).
 It is interesting to note that, when averages over all possible realizations of the q.d. are considered, 
classical Boltzmannian distributions still arise despite the absence of any evolution of the system at fixed boundary conditions:
 mapping the initial q.d. distribution over configurational distributions of damage states, via deterministic rules,
 still gives rise to classical statistical mechanics solutions. 

We have already shown the relationship between classical statistical mechanics and quasistatic fiber bundle models in the 
restricted framework of global sharing rule (Pride and Toussaint [15]), or in interacting fiber bundle models, with an energy based 
 evolution rule (Toussaint and Pride [16]). Here we show that the particular choice of energy 
or force based evolution rule does not alter the main results, and use a rule directly comparable to most numerical models.
\uneligne
{\center{\section{ MODEL DESCRIPTION AND EVOLUTION RULES }}}
We consider the following generic models: an ensemble of parallel fibers
are attached between parallel plates, at locations placed on the sites $\left\{x\right\}$ of a square lattice 
of dimensions $N=a \times a$, with the lattice step considered as length unit. The fibers are
supposed to have identical length $L_0$ at rest, and to
present the same spring constant, set to one which fixes the force
unit. A fiber thus carries a force $f=l$, up
to a threshold $t$ above which the fibers breaks irreversibly, so that afterwards $f=0$ independently of $l$. 
This threshold is picked ab initio independently for
each fiber, from a distribution $p(t)=dP(t)/dt$  where the cumulative
distribution $P(t)$ denotes the probability for a threshold to be below $t$.
From the initial rest state, the plates are separated while kept parallel,
and the minimum distance between them, $L_0+l_0$ , is increased in infinitesimal
steps. The lower plate is modelled as perfectly rigid, while the upper one
is allowed to have more complicated mechanical properties, which is
reflected in the load sharing rule between the fibers. When a fiber is broken at a location $x$, this creates a
force pertubation $-l_0$ on the corresponding site along the plates boundaries,
compared to the raw homogeneous force per site $l_0$ that would be exerted by a
bundle of intact fibers (counted positively in the direction from the
plates towards the bundle). This force perturbation possibly creates a
deformation of the plate boundary, illustrated in Fig. 1, if the plate is deformable,
\begin{equation} 
\delta l(y)=\varepsilon J(||y-x||) l_0 \label{force-perturb},
\end{equation}
where we have assumed isotropy and invariance of the system under
translation.
\begin{figure}
\centerline{\hbox{\includegraphics[%
  width=0.35\columnwidth,
  keepaspectratio]{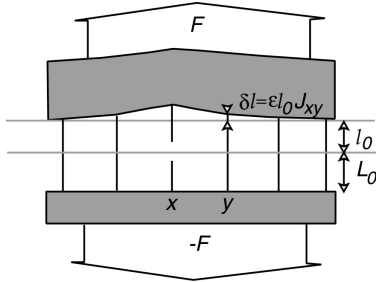}}}
\caption{Sketch of a fiber bundle between a rigid and a deformable plate, with the imposed raw elongation $l_0$ plus the elongation 
perturbation profile due to a broken fiber at location $x$.\label{bundle-sketch}}
\end{figure}
This is conceptually realized by considering biperiodic
boundary conditions along the lattice boundaries, with thresholding of the
interactions for separations exceeding the linear lattice dimension $a$. We
assume that $\varepsilon \ll 1$, and $0\leq J(r)\leq1$ for any separation $r$. 
The fact that $\delta l \ll l_0$ is 
granted if the plates are significantly stiffer than the fiber bundle, e.g. if they are rigid or elastic with a Young modulus much higher
than the fibers' spring constant divided by the elementary lattice site area.
The load sharing rule is then univoquely defined by the function $J(r)$. Classically considered cases are:
perfectly rigid plates (GLS), with $J(r)\equiv 0$ [8,9], or the opposite case (LLS),
where $J(r)=0$ for any separation $r$, except $J(r=1)=1$, i.e. interactions are only carried between nearest neighbors [10,11].
Power-law decay have also been considered [12] where $J(r)\sim r^{- \alpha}$,
the special case of $\alpha=1$ corresponding to a purely elastic plate [17]. Note also that an elastic sheet put under extension,
where circular flaws smaller than a lattice step nucleate when a threshold in local strain is
achieved, will present such power-law strain perturbations emanating from every flaw
with $\alpha=2$, which is thus a special case of the models discussed here.

The evolution rule of the system is as follows: for a certain quenched realization of the q.d. in the set of rupture thresholds$\left\{t_x\right\}$, the bundle is monotonically brought from zero to a macroscopic elongation $l_0$. Each time that the local elongation $l_0+\delta l(x)$ of some fiber at $x$ reaches its breaking threshold $t_x$, this fiber breaks irreversibly at $l_0$ kept fixed. The local elongations are then updated on all fibers according to Eq. (\ref{force-perturb}). If this leads to breakage of other fibers, the one corresponding to the minimum of $t_x/[l_0+\delta l(x)]$ is removed, and this avalanche procedure is iterated up to the point where all thresholds of surviving fibers are above their elongation. This completely defines a deterministic path for the state of the bundle as function of $l_0$ for each possible realization of the q.d.. The state of the bundle is referred to by an order parameter field $\varphi=\left\{\varphi_x\right\}$, with locally $\varphi_x=0$ for an intact fiber, and $\varphi_x=1$ for a broken one. For different realizations of the q.d., a priori different bundle states $\varphi$ will be obtained for identical final elongations $l_0$. For each possible state $\varphi$ and elongation, we define $P[\varphi,l_0]$ as the frequency of occurence of $\varphi$ at final elongation $l_0$, over all possible realizations of the initial q.d. -- i.e. when the deterministic experiment is performed ab initio as many times as there are possible q.d. realizations. We are interested in computing the configurational probability distribution $P[\cdot,l_0]$ for every elongation, and the associated average mechanical properties.

\uneligne
{\center{\section{RESULTS}}}

We first consider the simple Global Load Sharing rule.
Considering a fiber state $\varphi$, each fiber is brought at the same elongation $l_0$ independently of the remaining ones. Thus, the probability for this fiber to be broken is $P_0=P(l_0)$, and the probability that it has survived is $1-P_0$, independently of the other fibers. The state of each fiber at $l_0$ are then independent random variables, and the probability of a given state $\varphi$, which specifies the state of each fiber in every location, is simply
$P[\varphi,l_0]=P_0^n (1-P_0)^{N-n}$ where $n=\sum_x \varphi_x$ is the number of broken fibers in the state considered, and $N-n$ the number of surviving ones.
The configurational probability distribution is thus simply given by a site percolation model with probability of occupancy $P_0$.  

The average fraction of broken fibers at $l_0$ is thus $P_0$, and the mechanical properties are also directly obtained: Since the total force carried by the bundle is $F[\varphi,l_0]=l_0 \sum_x (1-\varphi_x)$, its average value over all realizations of the q.d. is $F=N l_0 (1-P_0)$, which is also the mechanical behavior of a fiber bundle in GLS with imposed total force (rather than elongation), in the limit of large sizes $N\rightarrow \infty$ [8,9]. If $F(l_0)$ presents a single maximum, this corresponds to a peak stress, which a priori does not coincide with the percolation transition happening at $P_0=1/2$. 
We notice that this configurational probability can also be cast under the form $P[\varphi,l_0]=\exp[-\ln[(1-P_0)/P_0] \sum_x \varphi_x] /Z$ where $Z$ is a normalization factor.

For more general Load Sharing Rules, i.e. for non-zero sharing functions $J$, the configurational probability distribution can also be obtained by perturbation, through the following reasoning: we consider a given state $\varphi$ and final elongation $l_0$. The elongation of each fiber in the final state, $l(x)=l_0[1+\varepsilon \sum_y J(||y-x||)]$, is by construction the maximum elongation that this fiber has reached from the initial state. We also have $l(x)\geq l_0$. We utilize these two facts to first obtain an upper bound of the probability $P[\varphi,x]$ via these arguments: if any fiber $x$ such as $\varphi_x=1$ in the considered $\varphi$ state had a threshold such as $t_x>l(x)$, this particular fiber should survive and the state $\varphi$ is not reached. Conversely, if there is any $x$ such as $\varphi_x=0$ in the considered state, having a threshold below $t_x<l(x)$, this fiber should break and the final state will also not be reached. Consequently, a necessary condition to reach the state $\varphi$ at $l_0$ is that each fiber such as $\varphi_x=1$ has a threshold realized below $l(x)$, and each one such as $\varphi_x=0$ had a threshold above $l(x)$. 
The probability of these two events to happen for each fiber threshold is $P(t<l(x))=l_0+l_0 \varepsilon \sum_y J(||y-x||))=P_0+ \delta P_x$, and $P(t>l(x))=1 - P_0 - \delta P_x$, where by Taylor expansion of the cumulative distribution $P$ around $l_0$ we define $\delta P_x=\varepsilon  l_0 \sum_y J(||y-x||)) p_0$, with $p_0=p(l_0)=dP/dx (l_0)$.
An upper bound to the probability of observing the considered state $\varphi$ is thus 
\begin{equation}
P[\varphi,l_0] \leq \prod_{\left\{x / \varphi_x =1\right\}} (P_0+ \delta P_x) \prod_{\left\{z / \varphi_z =0\right\}} (1 - P_0 - \delta P_z). \label{upper-bound}
\end{equation}

A lower bound can be obtained by expressing a set of sufficient conditions on the individual thresholds for the considered state $\varphi$ to be reached at $l_0$: if all fibers such as $\varphi_x=1$ had a threshold below $l_0$, while all the others had their thresholds above $l(x)$, the state $\varphi$ will be reached with certainty. In addition, other nonoverlapping subsets of the ensemble of realized thresholds lead with certainty to the considered state: if all fibers such as $\varphi_x=0$ have thresholds realized above $l(x)$, all fibers such as $\varphi_x=1$ but one have thresholds below $l_0$, and the last one has a threshold between $l_0$ and $l_0+\delta l(x)$, the first ones are intact with certainty, the second ones are broken with certainty under the effect of the basic mean field elongation, and thus the last considered fibers also breaks with certainty due to the elongation perturbations coming from the other broken ones. Adding the measures of these nonoverlapping subsets in the q.d. space (corresponding to the set of thresholds), we obtain a lower bound seeken for: 
\begin{equation}
P[\varphi,l_0] \geq \left\{ [\prod_{\left\{x / \varphi_x =1\right\}} (P_0)] + [\sum_ {\left\{x / \varphi_x =1\right\}} \delta P_x \prod_{\left\{y \neq x / \varphi_y =1\right\}} (P_0)] \right\} \prod_{\left\{x / \varphi_x =0\right\}} (1 - P_0 - \delta P_x).\label{lower-bound}
\end{equation}

To leading order $O(\varepsilon)$, these lower and upper bounds are identical, so that the above determines exactly the probability of occurence of configurations $\varphi$ that we look for, with interactions included through a perturbation analysis. This can be expressed more easily by considering the logarithm of the above: to leading order in the interactions $\varepsilon$, $P[\varphi,l_0]=e^{-H[\varphi,l_0]}/Z$ with Z a normalization factor, and
\begin{equation}
H[\varphi,l_0]=\ln\left( \frac{1-P_0}{P_0}\right) \sum_x \varphi_x - \frac{\varepsilon p_0 l_0}{P_0} \sum_{xy} J_{xy} \varphi_x \varphi_y + \frac{\varepsilon p_0 l_0}{1-P_0} \sum_{xy} J_{xy} \varphi_x (1-\varphi_y) \label{Hamiltonian}
\end{equation}
In this function, the first leading term corresponds to the mean field GLS result, and is analog to a chemical potential, influencing the average number of broken fibers at a given level. The second term reflects the tendency of damage to cluster due to stress perturbations, and is thus analog to a bulk energy. The third term arises from the fact that it is less likely for a fiber to be intact if it interacts with many broken ones. In the LLS model, this positive term would arise only at the boundary between broken and non broken clusters, and is thus analog to an interfacial tension.

It is possible to extract a formal temperature arising from all possible realizations of the initial q.d., directly related to the variance of elastic energy over all realized systems, by noting that the elastic energy in the ensemble of fibers is 
$E=(1/2) \sum_x (1-\varphi_x) (l_0+\delta l(x))^2 \simeq l_0^2/2 \sum_x (1-\varphi_x) + l_0^2 \varepsilon \sum_{xy} J(||y-x||) (1-\varphi_x) \varphi_y $, so that we can express $H[\varphi]=(E[\varphi]-\mu \sum_x \varphi_x)/T$ up to a constant, with a unique possible choice for the formal temperature and chemical potential, $T= 2 P_0(1-P_0)l_0/p_0$ and $\mu = 2 (P_0(1-P_0)l_0/p_0)\ln[P_0/(1-P_0)] - (l_0^2/2)[1+(4 P_0 - 1) \varepsilon \sum_x J(||x||)]$. If desired, it is then possible to use standard statistical mechanics techniques to derive from a potential defined as $-T \ln(Z)$, the statistical characteristics of the system as mechanical characteristics (sustained force by the bundle, averaged over all realizations of the q.d.), average number of broken fibers, Shannon entropy or autocorrelation function of the system [16]. 

Last, we note that Eq. (\ref{Hamiltonian}) maps these models onto well-known ones:
 defining a spin $\sigma_x=2 \varphi_x-1$, we can express $H=-\beta I \sum_x \sigma_x - \beta j J_{xy} \sum_{xy} \sigma_x \sigma_y$ with an external field $\beta I = \ln[(1-P_0)/P_0]/2 - (\varepsilon p_0 l_0 / 2 P_0) \sum_x J(||x||)$ and coupling factor $\beta j = \varepsilon p_0 l_0 / [4 P_0 (1-P_0)]$: this corresponds to generalized Ising models of coupling constant $j J_{xy}$, and reduces to standard Ising model for LLS, and percolation for GLS. Such models have a critical point at zero external field, and a certain value of coupling factor $\beta j_c$. When the external field $\beta I$ reverses sign, the coupling factor has a certain unique value $\beta j_r$, and such model should go through a percolation-like transition, if $\beta j_r \ll \beta j_c$, a first order transition if $\beta j_r>\beta j_c$, a second-order transition if $\beta j_r \sim \beta j_c$, or no transition if $0<\beta j_r<\beta j_c$.

\uneligne
{\center{\section{CONCLUSION}}}
For quasistatic interacting fiber bundle models with quenched disorder in the breaking thresholds, 
we have shown analytically how to obtain the probability distribution over damage configurations, 
when all possible realizations of the initial quenched disorder are considered. We have mapped these
q.d. based models onto paradigms of classical statistical physics, namely percolation, standard or generalized Ising models
for respectively, global, local or arbitrary decaying load sharing rules. The functional dependence of the coupling parameters 
over the elongation reached has been explicited analytically as forms which depend on the q.d. distribution and the load sharing rule considered.
This allows to obtain the possible phase transitions in such systems depending on these: second order ones associated to percolation,
 Ising or generalized Ising critical points, first order ones associated to (possibly generalized) Ising models, or none.
 This exact analytical mapping should be confronted to direct numerical testing in future work.

\uneligne
{\center{\section{REFERENCES}}}
[1] A.A. Griffith, {\em Philos. Trans. Roy. Soc. London A}, vol. 221, pp. 163 (1920).

[2] H.J. Herrmann and S. Roux, eds, {\em Statistical models for the fracture of disordered materials} (North-Holland, Amsterdam) (1990).

[3] A. Hansen and J. Schmittbuhl, Phys. Rev. Lett., vol. 90, 045504 (2003).

[4] E. Bouchaud, J. Phys. Condens. Matt. vol. 9, p.4319 (1997).

[5] J. Schmittbuhl and K.J. M{\aa}l{\o}y, Phys. Rev. Lett., vol. 78, p.3888 (1997).

[6] H.E. Daniels, Proc. Roy. Soc. A vol. 183, p.404 (1945).

[7] B.D. Coleman, J. Appl. Phys. vol. 29, p.968 (1958).

[8] D. Sornette, J. Phys. A vol. 22, p.L243 (1989).

[9] P.C. Hemmer and A. Hansen, Journal of Applied Mechanics, vol. 59, p.909 (1992).

[10] A. Hansen and P.C. Hemmer, Phys. Lettt. A, vol. 184, p.394 (1994).

[11] M. Kloster and A. Hansen and P.C. Hemmer, Phys. Rev. E, vol. 56, p.2615 (1997).

[12] R.C. Hidalgo, Y. Moreno, F. Kun and H.J. Herrmann, Phys. Rev. E, vol. 65, 046148 (2002).

[13] R.L. Blumberg-Selinger, Z.G. Wang, W.M. Gelbart and A. Ben-Shaul,  Phys. Rev. A, vol. 43, p.4396 (1991).

[14] A. Politi, S. Ciliberto and R. Scorretti, Phys. Rev. E, vol. 66, 026167 (2002).

[15] S.R. Pride and R. Toussaint, Physica A, vol. 312, p.159 (2002). 

[16] R. Toussaint and S.R. Pride, Cond-mat/0403412, preprint (2004).

[17]  J. Schmittbuhl, A. Hansen, and G. G. Batrouni, Phys. Rev. Lett. vol. 90, 045505 (2003).

\end{document}